 \documentclass{article}

\usepackage{amsmath}
\usepackage{amsfonts}
\usepackage{graphicx}
\usepackage{caption}
\usepackage{subcaption}

 \usepackage{amssymb}

\usepackage{epstopdf}
\usepackage{epsfig}
\usepackage{hyperref}

\newcommand{\insertplot}[5]{\begin{figure}
 \hfill\hbox to 0.05in{\vbox to #5in{\vfill
 \inputplot{#1}{#4}{#5}}\hfill}
 \hfill\vspace{-.1in}
 \caption{#2}\label{#3}
 \end{figure}}
 \newcommand{\inputplot}[3]{
 \special{ps: plotfile #1}
}
\newcounter{fig}

\newcommand{\ee}{\end{equation}}
\newcommand{\eea}{\end{eqnarray}}
\newcommand{\be}{\begin{equation}}
\newcommand{\bea}{\begin{eqnarray}}

\usepackage{graphicx}

\begin{document}

\title{  AdS$_5$ magnetized solutions in 
 minimal gauged supergravity }

\author{
{\large Jose Luis Bl\'azquez-Salcedo}$^{1}$,
{\large Jutta Kunz}$^{1}$,
{\large Francisco Navarro-L\'erida}$^{2}$
and
{\large Eugen Radu}$^{3}$
\vspace{0.5cm}
\\
$^{1}${\small Institut f\"ur  Physik, Universit\"at Oldenburg, Postfach 2503 
  D-26111 Oldenburg, Germany }
   \\
$^{2}${\small Dept.~de F\'{\i}sica At\'omica, Molecular y Nuclear, Ciencias F\'{\i}sicas 
 Universidad Complutense de Madrid, E-28040 Madrid, Spain}
   \\
$^{3}${\small Departamento de F\'\i sica da Universidade de Aveiro and CIDMA, Campus de Santiago, 3810-183 Aveiro, Portugal
 }
} 
 
\maketitle

\begin{abstract}   
 We construct a generalization of the AdS 
charged rotating black holes  
with two equal magnitude angular momenta
in five-dimensional minimal gauged supergravity.
In addition to the mass, electric charge and angular momentum,
the new solutions possess
an extra-parameter associated with a non-zero 
magnitude of the magnetic potential at infinity.
In contrast with the known cases, 
these new black holes possess a non-trivial zero-horizon size limit 
which describes a one parameter family of spinning charged solitons.  
All configurations reported in this work approach asymptotically 
an AdS$_5$ spacetime in global coordinates and are free of pathologies.
\end{abstract}

%\maketitle

%\vfill\eject
  \medskip
%%%%%%%%%%%%%%%%%%%%%%%%%%%%%%%%%%%%%%%%%%%%%%%%%%%%%%%%%%%%%%%%%%%
%\section{Introduction}
%%%%%%%%%%%%%%%%%%%%%%%%%%%%%%%%%%%%%%%%%%%%%%%%%%%%%%%%%%%%%%%%%%%
%%%%%%%%%%%%%%%%%%%%%%%%%%%%%%%%%%%%%%%%%%%%%%%%%%%%%%%%%%%%%%%%%%%%%%%%%%%%%%
\section{Introduction and motivation}
%%%%%%%%%%%%%%%%%%%%%%%%%%%%%%%%%%%%%%%%%%%%%%%%%%%%%%%%%%%%%%%%%%%%%%%%%%%%%%
The solutions of the five-dimensional gauged supergravity models
play a central role in the AdS/CFT correspondence
\cite{Maldacena:1997re},
\cite{Witten:1998qj}, 
providing a dual description of strongly-coupled conformal field theories (CFTs) on
the four-dimensional boundary of five-dimensional anti-de Sitter (AdS) spacetime.

In the minimal case, the bosonic sector  of the gauged supergravity model consists only of the graviton
and an Abelian vector field.  
However, despite its simplicity, constructing solutions of this theory is a nontrivial task, 
since the known generation  techniques do not work in the presence of a cosmological constant.
Thus one has to resort to trial and error or to numerical
calculations, starting from an appropriate Ansatz.
Restricting to stationary solutions approaching asymptotically a globally AdS$_5$ spacetime,
one notes that the problem greatly simplifies for the special case 
 where the two independent angular momenta
 of the generic configurations are set equal.
This factorizes the dependence on the angular coordinates, 
leading to a cohomogeneity-1 problem, with ordinary differential equations.
Subject to these assumptions, a general black hole (BH)
  solution has been found in closed form in  
	\cite{Cvetic:2004hs},
	\cite{Chong:2005hr}
	by Cveti\v c, L\"u and Pope (CLP).
This solution is characterized by three non-trivial parameters, namely the
mass, the electric charge, and one independent
angular momentum. 
These parameters are subject to some constraints, such that
closed timelike curves  
and naked singularities are avoided. 
{ Moreover, the CLP solution possesses an extremal limit
which preserves some amount of supersymmetry 
\cite{Gutowski:2004ez}.
} 
 
 A simple inspection of the  BH  in \cite{Cvetic:2004hs}
shows that 
it  does not possess a globally regular solitonic limit
which could be viewed as a deformation of the AdS background,
while the
magnetic field vanishes asymptotically. 
However, a number of recent studies
\cite{Herdeiro:2015vaa}-\cite{Herdeiro:2016plq}
have provided evidence that the previously known solutions
of the Einstein-Maxwell system in a globally AdS$_4$ background, represent only 
`the tip of the iceberg', being in some sense 
the AdS counterparts of the (well-known) Minkowski spacetime BHs.
A variety of
new configurations were shown to exist.
In strong contrast to the asymptotically flat case, 
this includes particle-like solitonic configurations
\cite{Herdeiro:2015vaa}, 
\cite{Costa:2015gol}
 and  even BHs with no spatial isometries \cite{Herdeiro:2016plq}.
Their existence can be traced back to the "box"-like behavior of
the AdS spacetime, which allows the existence of electric (or magnetic) multipoles, as test
fields, which are everywhere regular.

However, this "box"-like behavior is not specific to AdS$_4$ spacetime. 
It has been shown recently that cohomogeneity-1 solutions of Einstein-Maxwell theory in odd $D$ dimensions can be obtained with a non-vanishing magnetic field at the AdS$_D$ boundary\footnote{See also the more general  results in \cite{Chrusciel:2016cvr},
\cite{Chrusciel:2017emq}.
}  \cite{Blazquez-Salcedo:2016vwa}. These represent  new 
families of $static$ 
 solitons and black holes with rather different properties as compared to the well-known Reissner-Nordstr\"om-AdS solutions.
%which  
This result
suggests that
similar solutions should exist also for $D=5$ dimensions within the minimal gauged supergravity model. 
However, the Einstein-Maxwell-Chern-Simons case is more complex;
apart from the absence of the electric-magnetic duality,
we note that the solutions with a magnetic field necessarily rotate
and also that the sign of the electric charge becomes relevant \cite{Blazquez-Salcedo:2015kja,Blazquez-Salcedo:2016rkj}.

\medskip

This paper presents the results of a preliminary investigation in this direction,
by focusing on the simplest case of  configurations 
with equal
magnitude angular momenta.  
The new solutions reported here 
provide an extension of the CLP
BHs which contains
an additional parameter associated with the 
magnitude of the magnetic potential at infinity.
Our results show   the existence of a variety of new properties
of the solutions.
For example, the BHs possess a nontrivial
particle-like limit describing charged rotating solitons. 
Also, one finds solutions which rotate locally but have vanishing total angular momentum.

%%%%%%%%%%%%%%%%%%%%%%%%%%%%%%%%%%%%%%%%%%%%%%%%%%%%%%%%%%%%%%%%%%%%%%%%%%%%%%
\section{The model}
%%%%%%%%%%%%%%%%%%%%%%%%%%%%%%%%%%%%%%%%%%%%%%%%%%%%%%%%%%%%%%%%%%%%%%%%%%%%%% 

%%%%%%%%%%%%%%%%%%%%%%%%%%%%%%%%%%%%%%%%%%%%%%%%%%%%%%%%%%%%%%%%%%%%%%%%%%%%%%
\subsection{The action and equations }
%%%%%%%%%%%%%%%%%%%%%%%%%%%%%%%%%%%%%%%%%%%%%%%%%%%%%%%%%%%%%%%%%%%%%%%%%%%%%% 

The action for $D=5$ minimal
gauged supergravity  is given by 
\begin{equation} 
\label{EMCSac}
I= \frac{1}{16\pi } \int_{{\cal M}} d^5x\biggl[ 
\sqrt{-g}(R +\frac{12}{L^2} 
-F_{\mu \nu} F^{\mu \nu}  
+
\frac{2 \lambda }{3\sqrt{3}}\varepsilon^{\mu\nu\alpha\beta\gamma}A_{\mu}F_{\nu\alpha}F_{\beta\gamma}) 
 \biggr ] + I_{b},
\end{equation}
where $R$ is the curvature scalar, $L$ is the AdS length scale, 
$A_\mu $ is the gauge potential with the field strength tensor 
$ F_{\mu \nu} = \partial_\mu A_\nu -\partial_\nu A_\mu $ and $\varepsilon^{\mu\nu\alpha\beta\gamma}$ is the Levi-Civita tensor.
Also,
${ \lambda=1}$ is the Chern-Simons (CS) coupling constant.
However, $\lambda$ will be kept general 
in all relations below,
(such that (\ref{EMCSac})
will describe a generic Einstein--Maxwell--Chern-Simons (EMCS) model),
 although the numerical results will cover the SUGRA case only. 

In addition, (\ref{EMCSac}) contains a boundary term which is required for a consistent
variational principle and 
a proper renormalization of various physical quantities,
\begin{eqnarray}
\label{ct}
 I_{b}=-\frac{1}{8 \pi } \int_{\partial {\cal M}}d^{4}x\sqrt{-h}\biggl[
\rm{K}-\frac{3}{ L}(1+\frac{ L^2}{12}\rm{R})
-\frac{L}{2}
\log (\frac{L}{r}) \left \{{\rm F}_{ab}{\rm F}^{ab} \right \} 
\bigg]\ .
\end{eqnarray}
Here, 
$h_{ab}$ is the metric induced by $g_{\mu\nu}$ on the boundary
($\rm{R}$ being the corresponding Ricci scalar), 
and  $\rm{K}$ is the trace (with respect to
$h$) of the extrinsic curvature of the boundary.
Also, ${\rm F}_{ab}$ is the electromagnetic tensor 
induced on the boundary by the bulk field, 
while $r$ is a normal coordinate.

The field equations of this model consist of the Einstein equations
\begin{equation}
\label{Einstein_equation}
G_{\mu\nu} 
%+ \Lambda g_{\mu\nu} 
=\frac{6}{L^2}g_{\mu\nu}+2\left(F_{\mu\rho}{F^{\rho}}_{\nu}-\frac{1}{4}F^2 \right),
\end{equation}
together with the Maxwell--Chern-Simons   (MCS)  equations
\begin{equation}
\label{Maxwell_equation}
\nabla_{\nu} F^{\mu\nu} 
+ \frac{\lambda}{2\sqrt{3}}\varepsilon^{\mu\nu\alpha\beta\gamma}F_{\nu\alpha}F_{\beta\gamma}=0.
\end{equation}

%%%%%%%%%%%%%%%%%%%%%%%%%%%%%%%%%%%%%%%%%%%%%%%%%%%%%%%%%%%%%%%%%%%%%%%%%%%%%%
\subsection{The probe limit: Maxwell--Chern-Simons solutions in a fixed AdS background}
%%%%%%%%%%%%%%%%%%%%%%%%%%%%%%%%%%%%%%%%%%%%%%%%%%%%%%%%%%%%%%%%%%%%%%%%%%%%%% 
Before approaching the full model,
it is interesting to consider the probe limit, $i.e.$
a U(1) field in a fixed AdS spacetime with a line-element 
\begin{eqnarray}
\label{AdS}
&&ds^2 =
-N(r)dt^2+ 
\frac{dr^2}{N(r)}
  + \frac{1}{4}r^2  (\sigma_1^2+\sigma_2^2 + \sigma_3^2 )
,~~{\rm with}~~N(r)=1+\frac{r^2}{L^2}.
\end{eqnarray}
In the above line element, the (round) $S^3$ sphere is written 
 as an $S^1$-fibration over $S^2\equiv \mathbb C \mathbb P^1$, with
 $\sigma_i$ the left invariant one-forms, 
$
\sigma_1=\cos \psi d  \theta+\sin\psi \sin   \theta d \phi,
$
$
\sigma_2=-\sin \psi d \theta+\cos\psi \sin   \theta d \phi,
$
$
\sigma_3=d\psi  + \cos \theta d \phi;
$
%given by (\ref{sigma}).
also,  the coordinates $ \theta$, $\phi$, $\psi$ are the Euler angles on $S^3$,
with the usual range.
%
%%%%%%%%%%%%%%%%%%%%%%%%%%%%%%%%%%%%%%
\begin{figure}
    \centering
    
        \includegraphics[width=85mm,scale=0.75,angle=0]{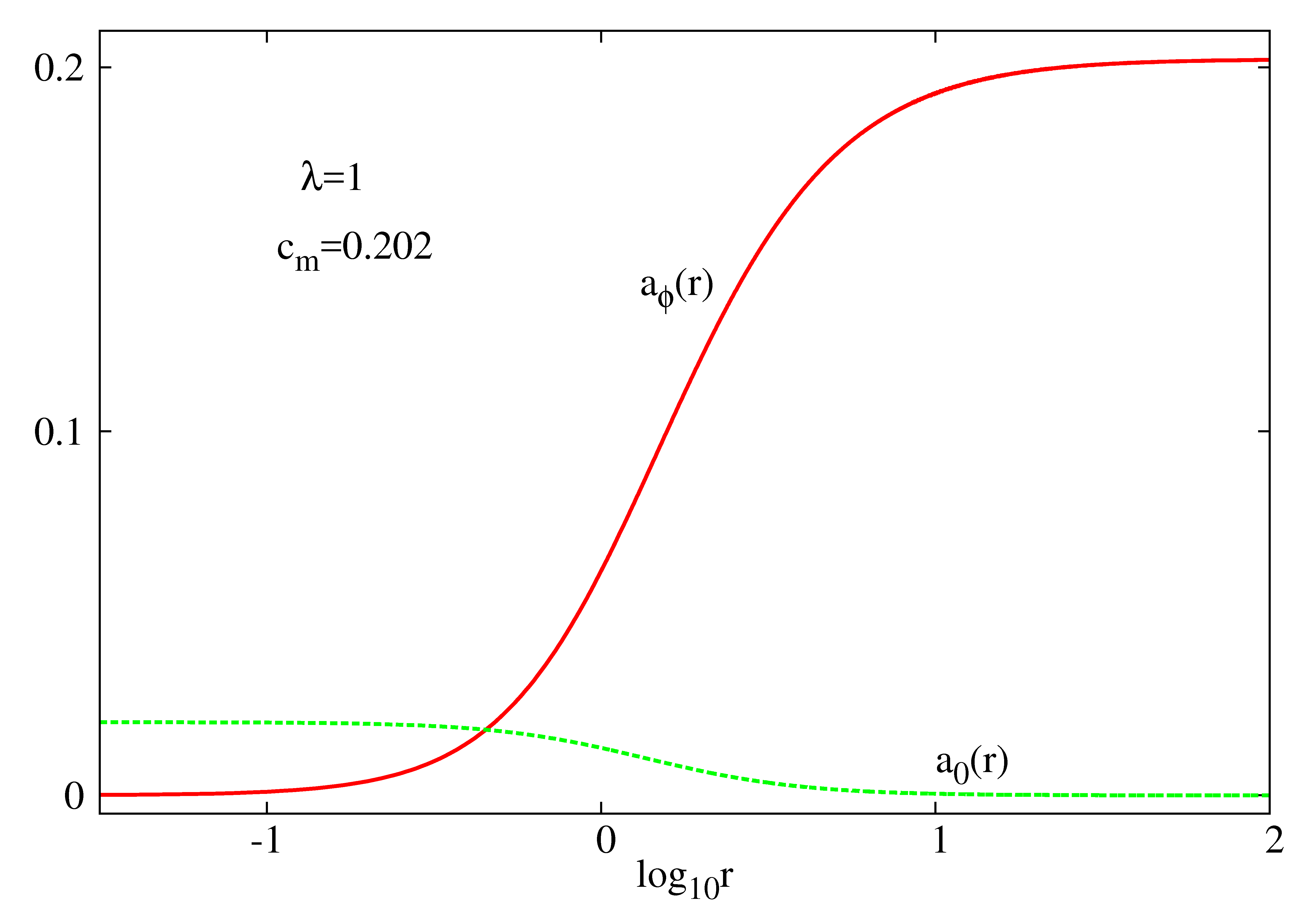}
        \caption{The profile of a typical solution of the Maxwell--Chern-Simons equations
in a fixed AdS background.}
        \label{soliton-profile}
\end{figure}

%%%%%%%%%%%%%%%%%%%%%%%%%%%%%%%%%%%%%%%%%%%%%%%%%%%%%%%%%%%%%%%%%%%%%%%%%%%%%%%%
%
The gauge field Ansatz contains an electric potential, $a_0$, and a magnetic one, $a_\varphi$ \cite{Cvetic:2004hs} 
\begin{eqnarray}
\label{U1}
A=a_0(r)dt +a_\varphi(r) \frac{1}{2} \sigma_3,
\end{eqnarray}
which results in the field strength tensor 
\begin{eqnarray}
\label{U1-F}
F=a_0'(r)dr\wedge dt +a_\varphi'(r) \frac{1}{2} dr\wedge \sigma_3+\frac{1}{2} a_\varphi(r)  \sigma_2\wedge \sigma_1~.
\end{eqnarray}
Then one writes the following  MCS  equations
\begin{eqnarray}
\label{probe1} 
\left(rN'a_\varphi' \right)'= \left(1+\frac{8}{3}\frac{\lambda^2}{r^2}a_\varphi^2 \right)\frac{4}{r}a_\varphi,
~~
 a_0'=-\frac{4\lambda a_\varphi^3}{\sqrt{3}r^3}~.
\end{eqnarray}
For $\lambda=0$ ($i.e.$ a pure Maxwell field in AdS spacetime)
the electric potential can be set to zero and
one finds the following exact solution
\begin{eqnarray}
\label{probe2} 
 a_\varphi(r)=c_m \left( 1-\frac{L^2}{r^2}\log(1+\frac{r^2}{L^2}) \right),
\end{eqnarray}
with $c_m$ an arbitrary (nonzero) constant.
Unfortunately, the Eqs. (\ref{probe1}) 
cannot be solved in closed form\footnote{Although one can construct a perturbative 
solution around (\ref{probe2}), this involves complicated special functions,
being not so useful.} for $\lambda \neq 0$.
However, such a solution exists; its small$-r$ expansion reads 
 (with $u,v_0,c_m$ and $\hat \mu$ nonzero constants):
\begin{eqnarray}
\label{probe3} 
a_\varphi(r)=u r^2+(-\frac{2}{3}\frac{u}{L^2}+\frac{8u^3\lambda^2}{9})r^4+\dots,~~
a_0(r)=v_0-\frac{2u\lambda}{\sqrt{3}}r^2+\dots,
\end{eqnarray}
while its form for $r\to \infty$ is
\begin{eqnarray}
\label{probe4} 
a_\varphi(r)=c_m+ \big(\hat\mu+2c_mL^2\log (\frac{L}{r}) \big)\frac{1}{r^2}+\dots,~~
a_0(r)=-\frac{2c_m^2 \lambda}{\sqrt{3}}\frac{1}{r^2}+\dots.
\end{eqnarray}
This implies the existence of a nonvanishing asymptotic magnetic field,
$F_{\theta \phi} \to -\frac{1}{2}c_m \sin\theta $, such that the parameter $c_m$
can be identified with the magnetic flux at infinity
through the base space $S^2$ of the $S^1$ fibration,
\begin{eqnarray}
\label{flux} 
\Phi_m=\frac{1}{4\pi}  \int_{S^2_\infty} F =-\frac{1}{2}c_m.
\end{eqnarray}
The smooth profiles connecting the asymptotics 
(\ref{probe3}) , 
(\ref{probe4}) 
are constructed numerically, a 
typical example being shown in Figure \ref{soliton-profile}.
%(note that all numerical results displayed in this work have $L=1$). 

Also, one can show that both $a_\varphi(r)$
and $a_0(r)$ are nodeless functions. 
Other properties of the MCS solutions 
are similar to those of their gravitating generalizations discussed in Section 3.
Moreover, rather similar configurations are found when considering instead
a Schwarzschild-AdS BH background.

%%%%%%%%%%%%%%%%%%%%%%%%%%%%%%%%%%%%%%%%%%%%%%%%%%%%%%%%%%%%%%%%%%%%%%%%%%%%%%
\subsection{The backreacting case}
%%%%%%%%%%%%%%%%%%%%%%%%%%%%%%%%%%%%%%%%%%%%%%%%%%%%%%%%%%%%%%%%%%%%%%%%%%%%%% 
When taking into account the backreaction on the geometry,
the solutions above should result in EMCS  solitons and BHs.  
Unfortunately, it seems that
no analytical techniques can be used  
to construct these solutions in closed form\footnote{
Some partial results can be found, however, in the
Einstein-Maxwell case ($\lambda=0$).
An approximate form of the static solitons ($a_0(r)=\omega(r)=0$)
can be constructed there
 by considering a perturbative expansion 
of the solutions 
in terms of the parameter $c_m$.  
}.
As such, in this work 
we approach this problem 
by solving the EMCS equations numerically,
subject to a
set of boundary conditions compatible with
an approximate expansion at the boundaries of the domain of integration\footnote{To integrate the
equations, we used the differential equation solver COLSYS which involves a Newton-Raphson method \cite{COLSYS}.
}.

The corresponding metric Ansatz is found by supplementing (\ref{AdS})
with four undetermined functions which take  into account the deformation of the AdS background
and factorize  the angular dependence
allowing for configurations with two equal angular momenta
\begin{eqnarray}
\label{metric}
&&ds^2 =
-f(r)N(r)dt^2+
\frac{1}{f(r)}
\left[
\frac{m(r)}{N(r)}dr^2
  + \frac{1}{4} r^2 \left(
	m(r) (\sigma_1^2+\sigma_2^2)+ n(r) \big(\sigma_3- \frac{2\omega(r)}{r} dt \big)^2
	\right)
	\right]
,
\end{eqnarray}
while the gauge field Ansatz\footnote{A rather similar framework 
has been used in 
\cite{Nedkova:2011aa} 
%\cite{Nedkova:2012yn}
to construct magnetized squashed BHs in $D=5$ Kaluza-Klein theory.
However, the properties of those solutions are very different.
}  
is still given by  (\ref{U1}).
This framework can be proven to be consistent, and, as a result, the EMCS equations reduce
to a set of six second order ordinary differential equations 
plus a first order constraint equation, whose expression can be found in Ref. \cite{Blazquez-Salcedo:2016rkj}.
Also, these equations possess two first integrals 
\begin{eqnarray} 
\label{FirstInt}
a_0'+\frac{\omega}{r}a_\varphi'- \frac{4\lambda}{\sqrt{3}}\frac{f^{3/2}a_{\varphi}^2}{r^3\sqrt{mn}}=
%{\frac {2\,{f}^{3/2}}{\sqrt {m}\sqrt {n}{r}^{3}\pi }}Q ,
{\frac {2\,{f}^{3/2}}{\pi  \sqrt {m n} {r}^{3}}}c_1,~~
~~~
%\begin{eqnarray}
%
 \frac{16\lambda}{3\sqrt{3}}a_{\varphi}^3-{\frac {{n}^{3/2}\sqrt {m}{r}^{3}}{{f}^{5/2}}}  
(r \omega' -\omega)
= c_2- {\frac {8\,c_1}{\pi }}\,a_{\varphi},
%=\frac{16}{\pi}J,
%\label{conJ}
\end{eqnarray}
where $c_1$ and $c_2$ are two constants.
 
%%%%%%%%%%%%%%%%%%%%%%%%%%%%%%%%%%%%%%%%%%%%%%%%%%%%%%%%%%%%%%%%%%%%%%%%%%%%%%
\subsection{The asymptotics}
%%%%%%%%%%%%%%%%%%%%%%%%%%%%%%%%%%%%%%%%%%%%%%%%%%%%%%%%%%%%%%%%%%%%%%%%%%%%%%  
%
In deriving the far field expression of the solutions, we impose that,  asymptotically,
$i)$ the geometry
becomes AdS in a static frame, 
$ii)$ the electric potential vanishes, $a_0\to 0$;
and, as a new feature as compared to the CLP case,
$iii)$ the magnetic potential  approaches a constant nonzero value, $a_\varphi\to c_m$.
Then a far-field expression of a solution  compatible 
with these assumptions  can be constructed
in a systematic way. 
The first few terms in this expansion read
\begin{eqnarray}
\nonumber
&&
f(r) = 1 + \left(\hat\alpha+\frac{12}{5}c_m^2L^2\log (\frac{L}{r}) \right) \frac{1}{r^4} +\dots, 
~
\omega(r) = \frac{\hat{J}}{r^3}-\frac{4q}{3}\left(\hat \mu-\frac{1}{3}c_m L^2(1-6\log(\frac{L}{r})) \right)\frac{1}{r^5}+\dots,
\\
\label{inf}
&&
m(r) = 1 + \left(\hat\beta+\frac{4}{5}c_m^2L^2\log (\frac{L}{r}) \right) \frac{1}{r^4} +\dots,
~
n(r) =1 + \left(3(\hat\alpha-\hat\beta)+\frac{4}{15}c_m^2L^2+\frac{24}{5}c_m^2L^2\log (\frac{L}{r}) \right) \frac{1}{r^4} +\dots,  
~~{~~~} 
\\
\nonumber
&&
a_{\varphi}(r) = c_m+ \left(\hat{\mu}+2c_m L^2 \log(\frac{L}{r}) \right)\frac{1}{r^2} +\dots,
~
a_{0}(r) = -\frac{q}{r^2} + \frac{c_m \lambda }{\sqrt{3}} \left( 2\hat \mu
+c_m L^2 \left(-1+4 \log(\frac{L}{r}) \right) \right)\frac{1}{r^4}+\dots, 
\end{eqnarray}
%This asymptotic expansion contains 6 undetermined parameters
%\begin{eqnarray}
with
$ \{
\hat\alpha,\hat\beta, \hat J; c_m, \hat \mu,q 
 \} 
$ 
undetermined
parameters. 
%Note that these expressions are compatible with the probe limit expansion (\ref{probe4}) up to order $O(c_m^2)$.

%%%%%%%%%%%%%%%%%%%%%%%%%%%%%%%%%%%%%%%%%%%%%%%%%%%%%%%%%%%%%%%%%%%%%%%%%%%
%\subsubsection{Solitons: small-$r$ expansion}
%%%%%%%%%%%%%%%%%%%%%%%%%%%%%%%%%%%%%%%%%%%%%%%%%%%%%%%%%%%%%%%%%%%%%%%%%%%
Concerning the solitons,
one can also construct a  small-$r$ approximate form of the solutions
%The expression of the solutions as $r\to 0$ can be written 
as a power series in $r$, 
compatible with the assumption of regularity at $r=0$.
The first terms in this expansion 
are\footnote{ {It is interesting to contrast these asymptotics with those satisfied by the topological
solitons in \cite{Cvetic:2005zi} 
which, however, possess a vanishing magnetic field at infinity, $c_m=0$.
For topological solitons, 
the proper size of the $\psi$-circle goes to zero as $r\to 0$,
while the coefficient of the round $S^2$-part in (\ref{metric}) is positive
(this holds also for $g_{rr}$ and $-g_{tt}$).}
}
\begin{eqnarray}
\nonumber
&&
f(r) = f_0 + \left(\frac{m_0-f_0}{L^2}+\frac{4 u^2 f_0^2}{3m_0} \right)r^2 +\dots, 
~~
m(r) =m_0+m_2 r^2 +\dots, ~~
\omega(r)=w_1 r-\frac{8u^3f_0^{5/2}\lambda}{3\sqrt{3}m_0^2}r^2+\dots,
\\
\label{zero}
&&
n(r) = m_0+ \left(\frac{3m_0(m_0-f_0)}{f_0 L^2}-m_2+\frac{4 u^2 f_0}{3} \right)r^2 +\dots,~~
a_{0}(r) =  v_0-\left(\frac{2u^2 f_0^{3/2}\lambda}{\sqrt{3} m_0}+u w_1 \right)r^2+\dots,  
\\
\nonumber
&&
a_{\varphi}(r) = u r^2+ \frac{u}{9f_0L^2 m_0}\bigg( 4u^2f_0^2L^2(1+2\lambda^2)+3(4m_0^2-3f_0(2m_0+L^2M_2)) \bigg)r^4+\dots,
\end{eqnarray}
with the free parameters
%\begin{eqnarray}
 $
\{
 f_0, m_0,m_2, w_1; u,v_0
  \}.
	$
 
%%%%%%%%%%%%%%%%%%%%%%%%%%%%%%%%%%%%%%%%%%%%%%%%%%%%%
 %\subsubsection{Black holes: near horizon expansion}
%%%%%%%%%%%%%%%%%%%%%%%%%%%%%%%%%%%%%%%%%%%%%%%%%%%%%
However, when gravitating
solitons exist in a given model, 
normally
one can also construct
bound states of such solitons with an event horizon
\cite{Kastor:1992qy},
\cite{Volkov:1998cc}.
These BHs have a horizon which 
  is a squashed $S^3$ sphere and
resides at a constant value of the quasi-isotropic radial coordinate $r = r_H>0$.
The non-extremal solutions\footnote{We have found numerical evidence 
for the  existence of extremal BHs as well.
Such solutions possess a  different near horizon expression, while
 the far field expansion (\ref{inf}) holds also in that case.
The extremal BHs possess a number of distinct features and will be reported elsewhere  
(however, some properties of the $T_H=0$ limit can be seen in Figure 5).
} 
have the following expansion valid as $r\to r_H$ 
\begin{eqnarray}
\nonumber
&&
f(r) = f_2 (r-r_H)^2+ O\left(r-r_H\right)^3,~~ 
m(r)= m_2 (r-r_H)^2 + O\left(r-r_H\right)^3, 
\\
\label{near-horizon}
&&
n(r) = n_2 (r-r_H)^2  
+ O\left(r-r_H\right)^3, ~~
  \omega (r) = \omega_0  +  O\left(r-r_H\right), 
\\
	\nonumber
&&
a_{0}(r) =a_{0 }^{(0)} + O\left(r-r_H\right)^2,
~~
a_{\varphi}(r) =  a_{\varphi }^{(0)}+ O\left(r-r_H\right)^2,
\end{eqnarray}
with 
 $
\{
f_2,m_2,n_2,\omega_0;a_{0 }^{(0)} ,a_{\varphi }^{(0)}
\}$
undetermined parameters.
 Also, note that the behavior of solutions inside the horizon ($r<r_H$)
is not discussed in this work.

%%%%%%%%%%%%%%%%%%%%%%%%%%%%%%%%%%%%%%%%%%%%%%%%%%%%%
\subsection{Physical parameters}
%%%%%%%%%%%%%%%%%%%%%%%%%%%%%%%%%%%%%%%%%%%%%%%%%%%%%

In the next Section we give numerical evidence for the
existence of smooth EMCS solutions interpolating between
the asymptotics above.
Most of the physical properties
can be read off from the asymptotic
data near the horizon/origin and at infinity.
 
{
The mass $M$ and angular momentum $J$ of these solutions is computed by using
the quasilocal formalism 
\cite{Balasubramanian:1999re},
with a boundary stress tensor ${\rm T}_{ab}=\frac{2}{\sqrt{-h}}\frac{\delta I}{\delta h^{ab}}$.
Then $M$ and  $J$ are the conserved charges associated with Killing
symmetries 
 $ \partial_t$, 
 $ \partial_\psi$
of the induced boundary metric $h$, found for a large constant value  of $r$. 
}
This results in\footnote{Note that 
$M$ and $J$ 
are evaluated relative to a frame which is  nonrotating  at infinity.
%Also, the boundary stress tensor acquires a nonzero contribution from 
%magnetic field on the boundary
}
\begin{eqnarray}
M = -\frac{\pi}{8}\frac{(3\hat\alpha+\hat\beta)}{L^2} +\frac{c_m^2 \pi}{30}+\frac{3\pi}{32}L^2,~~~~ 
J = \frac{\pi}{4}\hat{J}.
\end{eqnarray} 
The  electric charge $Q$, as computed from the usual definition, is
\begin{equation}
Q= - \frac{1}{2} \int_{S_{\infty}^{3}}
 %\left(
 \tilde F 
%+\frac{\lambda}{\sqrt{3}} A \wedge F
% \right)
%= - \frac{1}{2} \int_{S_{\infty}^{3}} \tilde F 
= \pi q~,
\label{charge}
\end{equation}
with 
${\tilde F}_{\mu_1 \mu_2 \mu_3} \equiv  
  \epsilon_{\mu_1 \mu_2 \mu_3 \rho \sigma} F^{\rho \sigma}$.
% and $\epsilon_{\mu_1 \mu_2 \mu_3 \rho \sigma}$ the Levi-Civita tensor. (no need-- was introduced already).
However, this quantity is not related to any conservation law if $c_m \neq 0$.
 A more appropriate definition is now the Page charge
\cite{Page:1984qv},  
\cite{Marolf:2000cb},
\begin{equation}
Q^{(P)}= - \frac{1}{2} \int_{S_{\infty}^{3}} \left( \tilde F 
+\frac{\lambda}{\sqrt{3}} A \wedge F \right)
%= - \frac{1}{2} \int_{S_{\infty}^{3}} \tilde F 
= Q - \frac{2\pi}{\sqrt{3}}\lambda c_m^2 \equiv c_1~,
\label{Pcharge}
\end{equation}
being related to the total derivative structure of the Maxwell-Chern-Simons equations
(with $c_1$ the integration parameter we introduced in the first integral (\ref{FirstInt})).

Another physically relevant parameter one can define is the R-charge, 
associated with the conservation of the R-current of the dual theory at the AdS boundary \cite{Aharony:1999ti}:
\begin{equation}
Q^{(R)}= - \frac{1}{2} \int_{S_{\infty}^{3}} \left( \tilde F 
+\frac{2\lambda}{3\sqrt{3}} A \wedge F \right)
%= - \frac{1}{2} \int_{S_{\infty}^{3}} \tilde F 
= Q - \frac{4\pi}{3\sqrt{3}}\lambda c_m^2~.
\label{Rcharge}
\end{equation} 
Note that these three charges coincide in the absence of a boundary magnetic field, $c_m=0$.
Also,  the first integral (\ref{FirstInt}) 
implies that  $ \frac{1}{2}c_m Q^{(R)}+J=-\frac{\pi}{16}c_2$.

In the above relations, $\alpha$, $\beta$, $\hat J$ and $q$
are parameters which enter the far field expansion  (\ref{inf}). 
Also, we remark that the interpretation proposed  for
$c_m$ 
in the probe limit,
as a magnetic flux at infinity,
still holds in the backreacting case.

%In addition the integration parameter $c_2$ from equation (\ref{FirstInt}) is related to the total angular momentum and magnetic flux of the %system:
%\begin{equation}
%\frac{\pi}{16}c_2=-\frac{1}{2}c_m Q^{(R)}-J= \Phi_m Q^{(R)}-J.
%\label{c2_exp}
%\end{equation}
 
%%%%%%%%%%%%%%%%%%%%%%%%%%%%%%%%%%%%%%%%%%%%%%%%%%%%%
%\subsubsection{Horizon quantities}
%%%%%%%%%%%%%%%%%%%%%%%%%%%%%%%%%%%%%%%%%%%%%%%%%%%%%
Turning now to BH quantities defined in terms 
of the horizon boundary data in  (\ref{zero}), we note that
the solutions' horizon angular velocity is
\begin{equation} 
\Omega_H=\frac{\omega_0}{r_H},
\end{equation}
while
the area of the horizon $A_H$ and the Hawking temperature $T_H$ of the solutions are given by
\begin{equation} 
A_H= 
2\pi^2r_H^3 \frac{m_2}{f_2}\sqrt{\frac{n_2}{f_2}},~~~~
T_H= \frac{1}{2\pi}\left(1+\frac{r_H^2}{L^2} \right)\frac{f_2}{\sqrt{m_2}}~.
\end{equation}   
The horizon electrostatic potential $\Phi_H$ 
as measured in a co-rotating frame on the horizon is
%is defined by
%(with $\zeta = \partial_t + \Omega_H \partial_\psi$) 
\begin{equation}
\Phi_H 
% \left. \zeta^\mu A_\mu \right|_{r=r_{\rm H}} 
= a_{0 }^{(0)}+\Omega_H a_{\varphi}^{(0)}.
\label{Phi} 
\end{equation} 
Also, to have a measure of the squashing of the horizon, 
we introduce the deformation parameter 
\begin{eqnarray}
\varepsilon=\frac{n(r)}{m(r)}\bigg|_{r=r_H}=\frac{n_2}{m_2},
\end{eqnarray} 
which gives the ratio of  the $S^1$ and the round $S^2$  
parts of the (squashed $S^3$) horizon metric, respectively.

Finally, let us remark that all configurations reported in this work have $f,m,n$ 
strictly positive functions for $r>r_H$ (or $r\geq 0$ for solitons).
As such, $t$ is a global time coordinate and the metric 
is free of causal pathologies \cite{Cvetic:2005zi}.
We have also monitored the Ricci and the Kretschmann scalars 
of the solutions
and did not find any indication for a singular behavior.

%%%%%%%%%%%%%%%%%%%%%%%%%%%%%%%%%%%%%%%%%%%%%%%%%%%%%%%%%%%%%%%%%%%%%%%%%%%%%%
\section{The solutions}
%%%%%%%%%%%%%%%%%%%%%%%%%%%%%%%%%%%%%%%%%%%%%%%%%%%%%%%%%%%%%%%%%%%%%%%%%%%%%%
%%%%%%%%%%%%%%%%%%%%%%%%%%%%%%%%%%%%%%%%%%%%%%%%%%%%%%%%%%%%%%%%%%%%%%%%%%%%%%
\subsection{Solitons}
\begin{figure}
    \centering
		
    \begin{subfigure}[b]{0.4\textwidth}
        \includegraphics[width=75mm,scale=0.75,angle=-0]{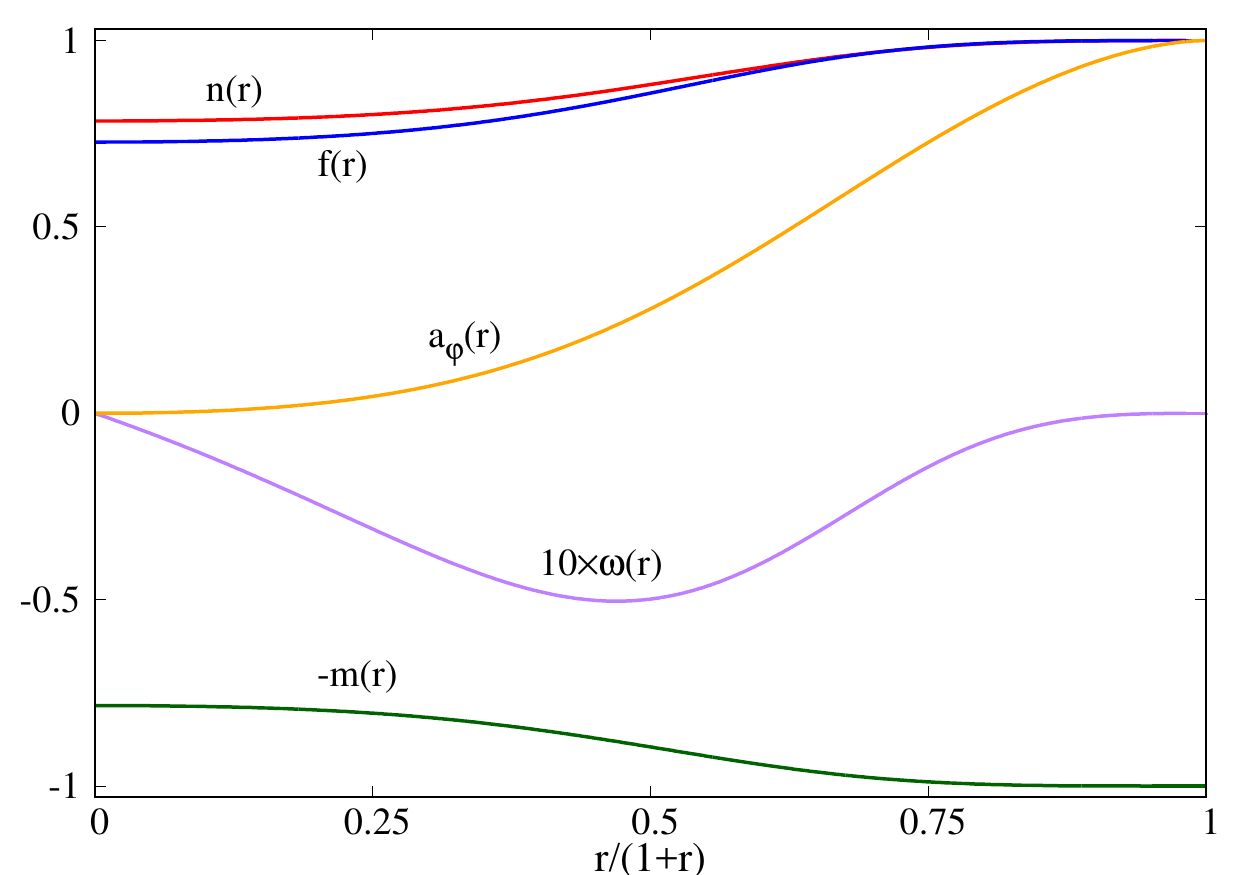}
        %\label{fig:ak_excitations}
    \end{subfigure}~~~~
    \begin{subfigure}[b]{0.4\textwidth}
        \includegraphics[width=75mm,scale=0.75,angle=-0]{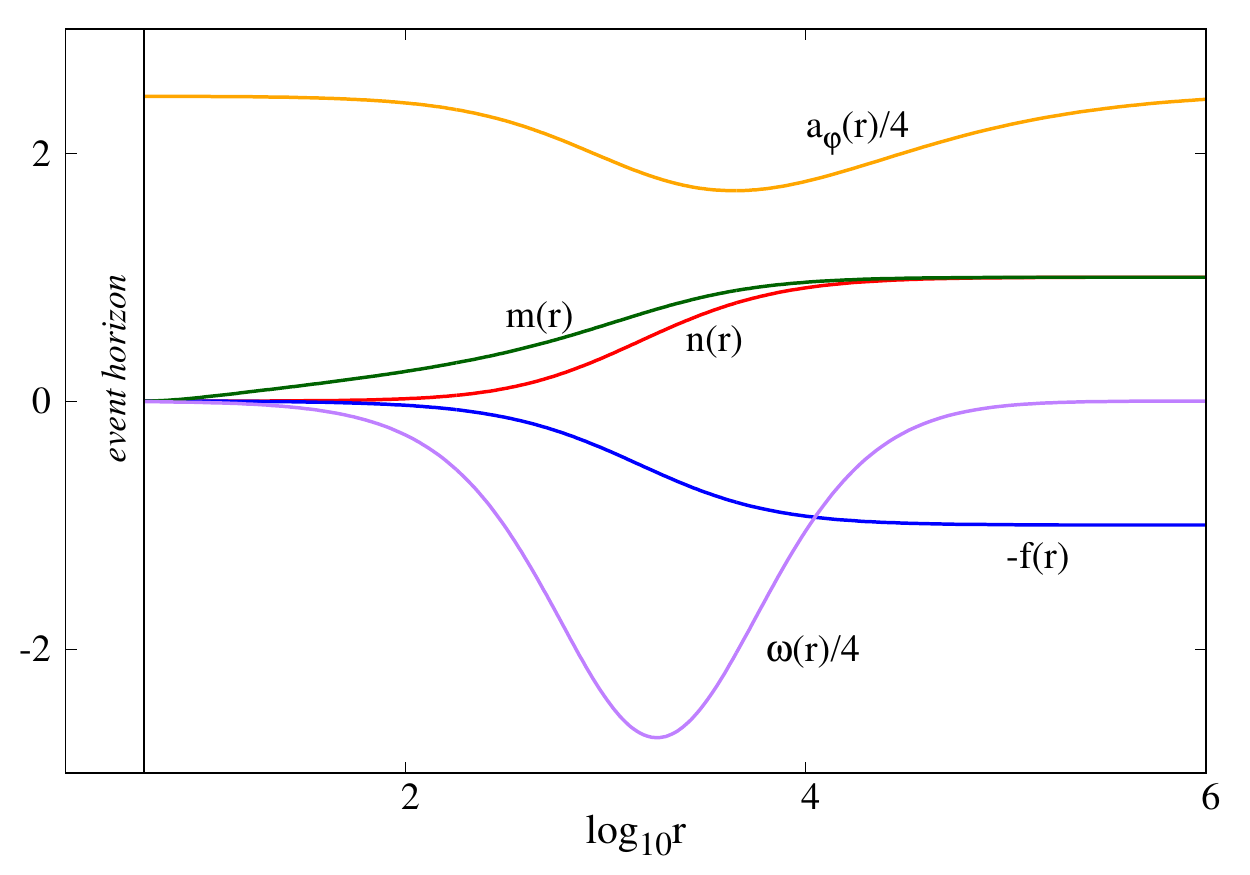}
        %\label{fig:omega_excitations}
    \end{subfigure}
 
		    \caption{The profiles of a typical soliton with $c_m=1$, $L=1$ (left) and a typical black hole with $r_H=2$, $J=0$, $Q=10$, $c_m=10$ and $L=15$ (right).}
		\label{profiles}		
\end{figure}

%%%%%%%%%%%%%%%%%%%%%%%%%%%%%%%%%%%%%%%%%%%%%%%%%%%%%%%%%%%%%%%%%%%%%%%%%%%%%%
 The numerical results indicate the existence of 
a family of everywhere regular solutions with
finite mass, charge and angular momentum.
Such configurations can be viewed as deformations of the (globally) AdS background,
corresponding  to charged, spinning EMCS solitons.
{ They possess no horizon, while the size of both parts of the $S^3$-sector of the metric shrinks to zero as $r\to 0$.}
The profile of a typical solution is exhibited in Figure \ref{profiles} (left).

%%%%%%%%%%%%%%%%%%%%%%%%%%%%%%%%%%%%%%%%%%%%%%
%%%%%%%%%%%%%%%%%%%%%%%%%%%%%%%%%%%%%%

The solitons have rather special properties.
The only input parameter here is the constant $c_m$
which fixes the magnitude at infinity of the magnetic  potential.
For any $\lambda\neq 0$,
the electric charge and angular momentum are given by\footnote{The  
relations (\ref{rel1}) are found
by evaluating the first integrals (\ref{FirstInt}) for 
  the asymptotic expansions (\ref{inf}), (\ref{zero}) (note that the solitons have $c_1=c_2=0$).}
\begin{eqnarray}
\label{rel1}
J= -\frac{ \lambda \pi}{3\sqrt{3}}c_m^3,~~~~Q=3Q^{(R)}=\frac{2\lambda \pi }{\sqrt{3}}c_m^2 ~,
\end{eqnarray}
such that the following  universal relation is satisfied
\begin{eqnarray}
%J=-\frac{1}{6}c_m Q= \frac{1}{3}\Phi_m Q,
J=
%-\frac{1}{2}c_m Q^{(R)}=
 \Phi_m Q^{(R)},
\end{eqnarray}
with $\Phi_m $  computed from (\ref{flux}).

\begin{figure}
    \centering
    
        \includegraphics[width=75mm,scale=0.75,angle=0]{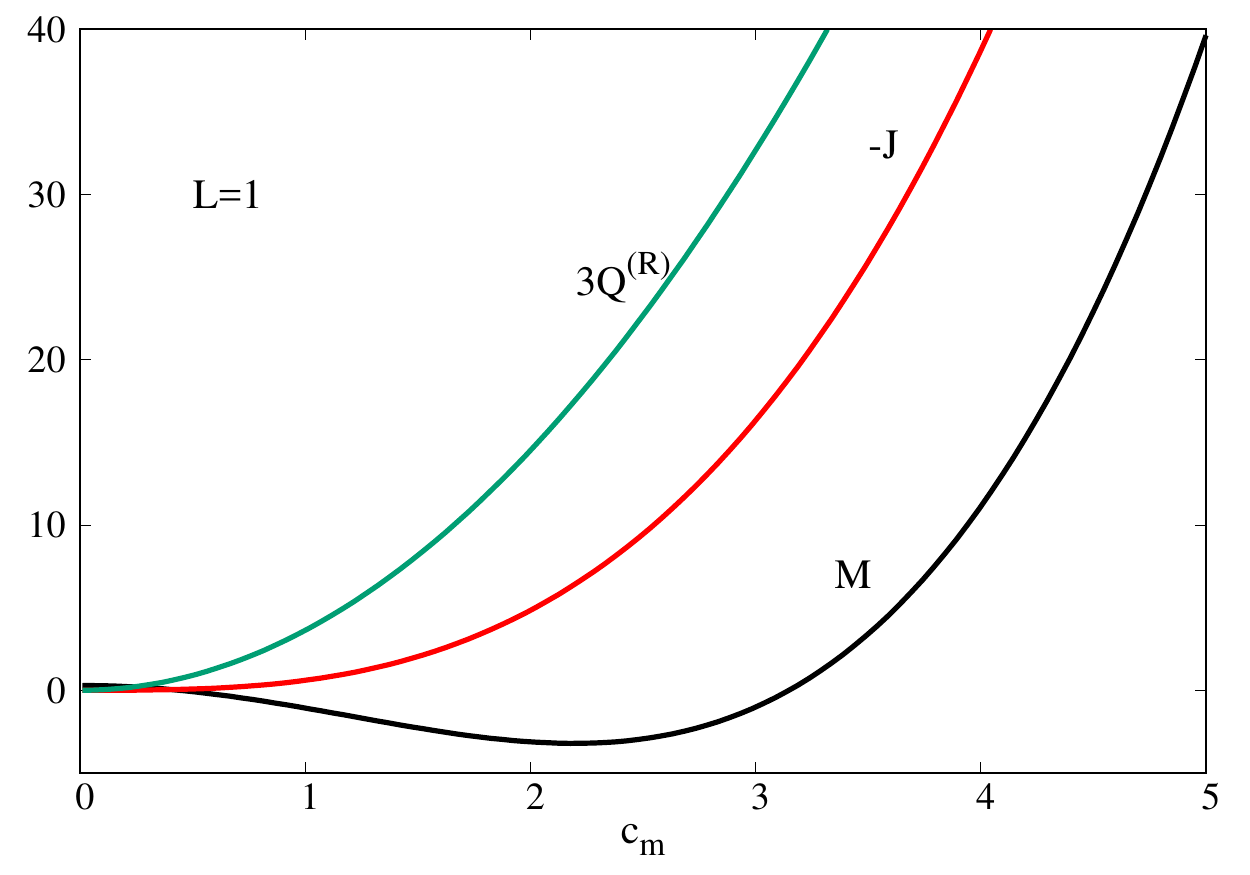}
        \caption{The mass, angular momentum and electric charge of 
gravitating  spinning solitons  are shown as a function of the  magnetic parameter $c_m$. 
}
        \label{soliton-mass}
				
\end{figure}

The $M(c_m)$ dependence can be found only numerically, being displayed in Figure \ref{soliton-mass}.
A good fit up to a relatively high value $c_m\sim 4$ reads
\begin{eqnarray}
 \frac{M}{L^2} =\frac{3\pi}{32} + a_2 \frac{c_m^2}{L^2} + a_4 \frac{c_m^4}{L^4}  + a_6  \frac{c_m^6}{L^6}~, 
\end{eqnarray}
 with 
$a_2= -1.52$,  
$a_4= 0.175$,  
$a_6= -0.0025$,
and variance of residuals of $3 \times 10^{-4}$.
Note that no upper bound on $|c_m|$ seems to exist; 
however, the numerics becomes difficult for large values of it.  
Also, similar to the probe limit, we could not find excited solutions 
(which would possess a magnetic potential with nodes \cite{Blazquez-Salcedo:2016rkj});
this holds also in the BH case.
However, we conjecture the existence of such solutions 
for  large enough values of the CS coupling constant $\lambda$.

 %%%%%%%%%%%%%%%%%%%%%%%%%%%%%%%%%%%%%%%%%%%%%%%%%%%%%%%%%%%%%%%%%%%%%%%%%%%%%%
\subsection{Black holes}
%%%%%%%%%%%%%%%%%%%%%%%%%%%%%%%%%%%%%%%%%%%%%%%%%%%%%%%%%%%%%%%%%%%%%%%%%%%%%%
As expected, these solutions possess BH generalizations.
They can be constructed starting with $any$
CLP solution and slowly increasing the value of the parameter $c_m$.
The profile of a typical BH is shown in Figure \ref{profiles} (right).

Finding the domain of existence of these BHs 
together with their general properties
is a considerable task
which is not aimed at in this paper. 
Instead, we analyze several particular classes  of solutions, 
looking for special properties.
 
 %%%%%%%%%%%%%%%%%%%%%%%%%%%%%%%%%%%%%%
%
%%%%%%%%%%%%%%%%%%%%%%%%%%%%%%%%%%%%%%%%%%%%%%%%%%%%%%%%%%%%%%%%%%%%%%%%%%%%%%%%

\begin{figure}
    \centering
		
    \begin{subfigure}[b]{0.42\textwidth}
        \includegraphics[width=50mm,scale=0.5,angle=-90]{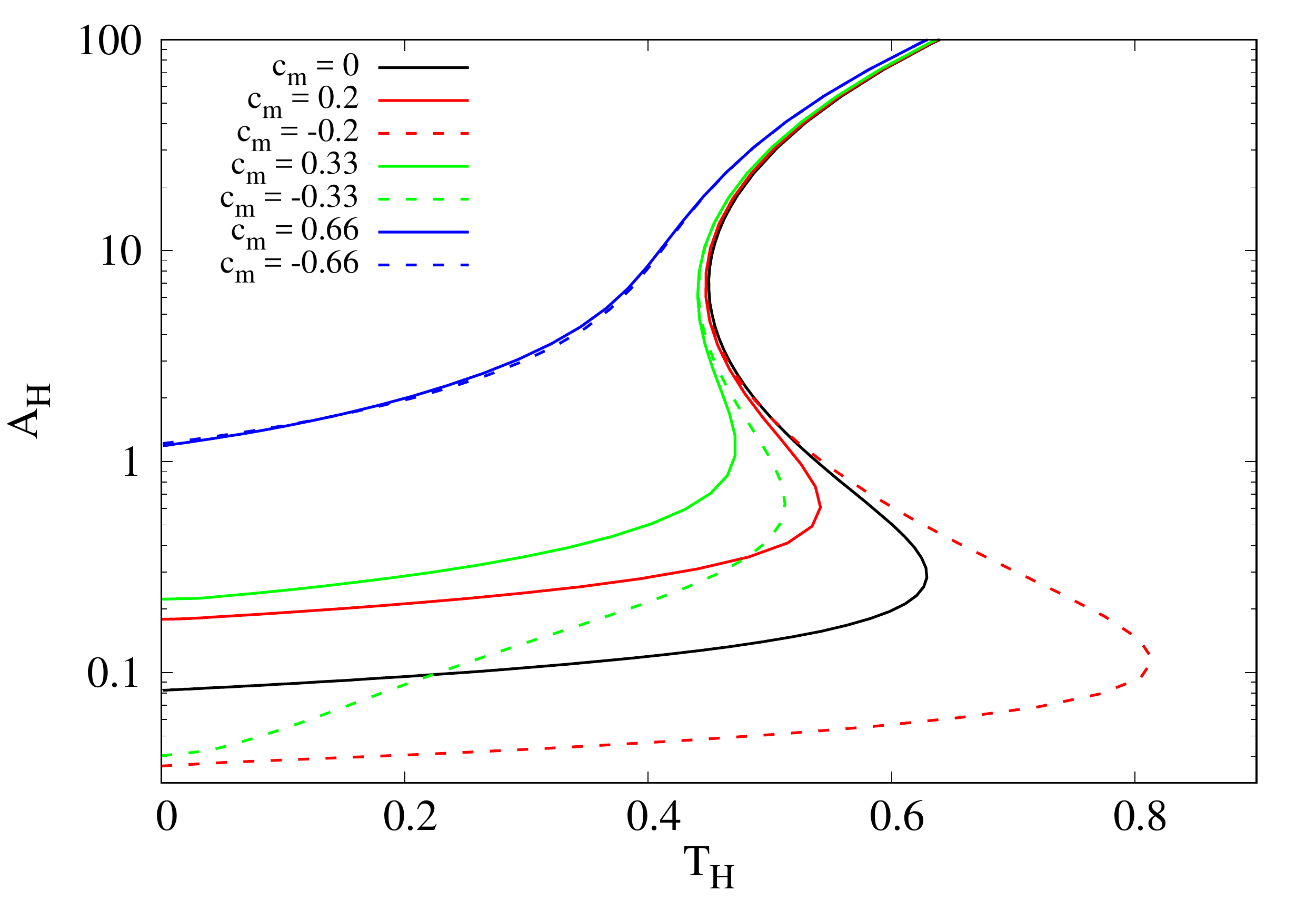}
        %\label{fig:ak_excitations}
    \end{subfigure}~~~~
    \begin{subfigure}[b]{0.42\textwidth}
        \includegraphics[width=50mm,scale=0.5,angle=-90]{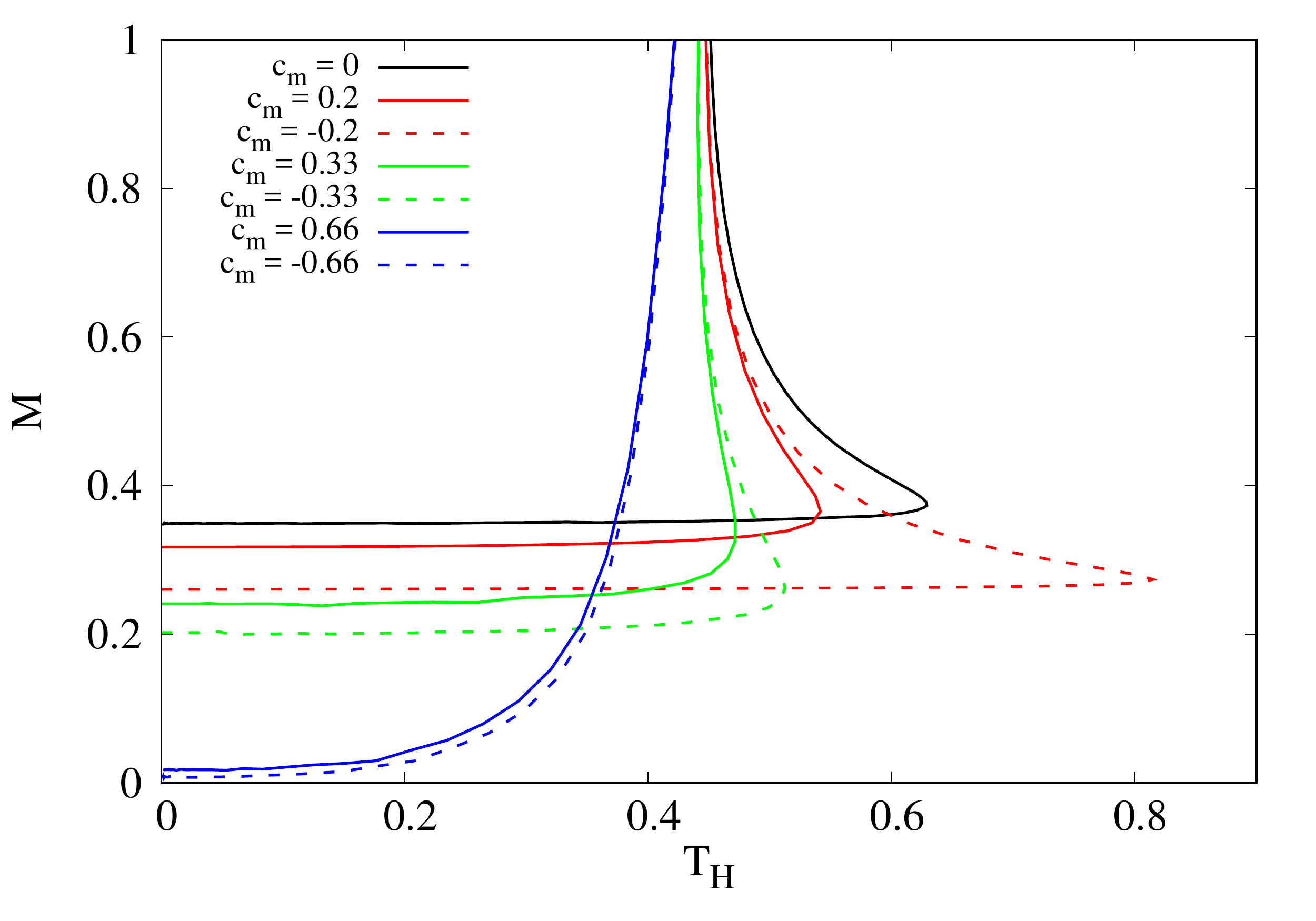}
        %\label{fig:omega_excitations}
    \end{subfigure}
		\\
     \begin{subfigure}[b]{0.42\textwidth}
        \includegraphics[width=50mm,scale=0.5,angle=-90]{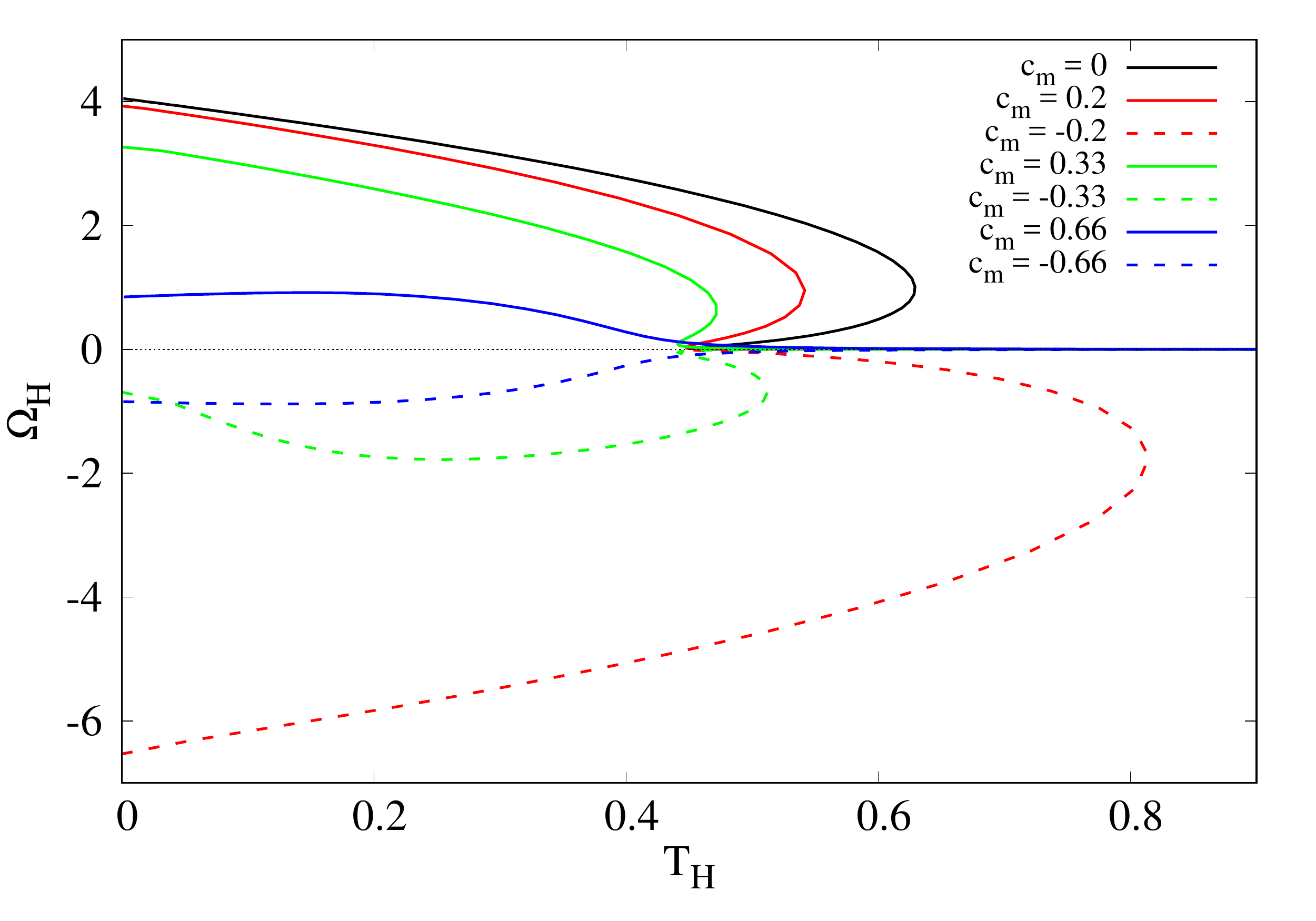}
        %\label{fig:ak_excitations}
    \end{subfigure}~~~~
    \begin{subfigure}[b]{0.42\textwidth}
        \includegraphics[width=50mm,scale=0.5,angle=-90]{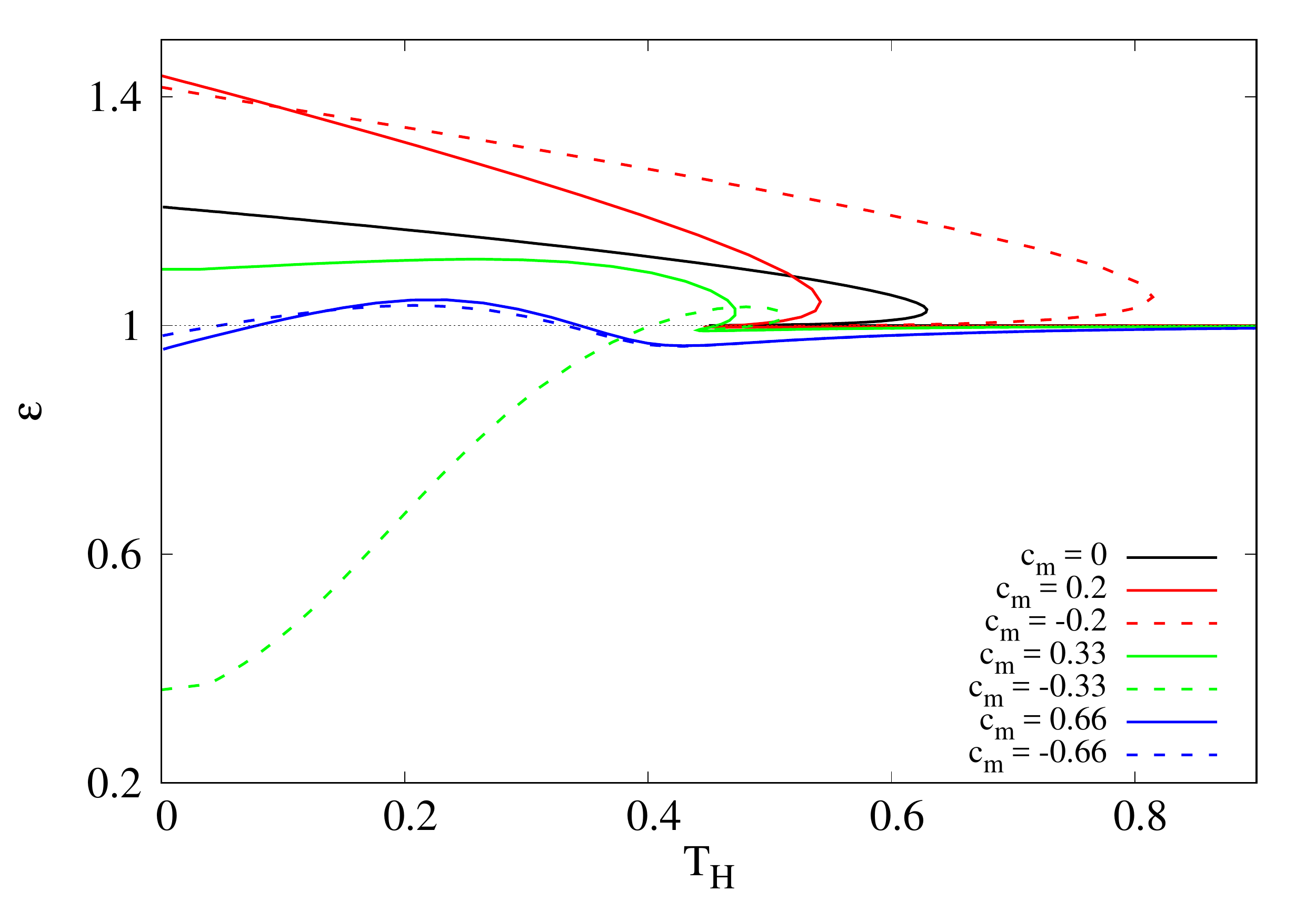}
    \end{subfigure}
		    \caption{Area $A_H$, mass $M$, angular velocity $\Omega_H$ and horizon deformation $\varepsilon$, as a function of the horizon temperature $T_H$ and for 
 several values of the magnetic flux on the boundary.
%These solutions have $L=1$, an electric charge $Q=0.0044$ and an angular momentum $J=0.003$.}
These solutions have AdS length $L=1$, angular momentum $J=0.003$, and the R-charge $Q^{(R)}=0.044$.}
		\label{BH-prop}		
\end{figure}

In Figure \ref{BH-prop} we display the results  
for configurations with fixed values of 
both $Q^{(R)}$ and $J$
and several values of $c_m$.
The first feature we notice is that 
$c_m\neq 0$
leads to some differences for small values of $T_H$ only,
while the solutions with large temperatures are essentially CLP BHs.
Also, as expected,  
the qualitative behavior of solutions with small $|c_m|$ resembles that of the unmagnetized case.
However, this changes for large enough values of  $c_m$
and one finds $e.g.$
a monotonic behavior of mass and horizon area as a function of temperature.
In particular, this means that for large values of $|c_m|$, the BHs become thermodynamically stable
for the full range of $T_H$,
with the existence of one branch of solutions only.
Also, the sign of $c_m$ is relevant for small values of $|c_m|$, only.

%%%%%%%%%%%%%%%%%%%%%%%%%%%%%%%%%%%%%%%%%%%%%%
%%%%%%%%%%%%%%%%%%%%%%%%%%%%%%%%%%%%%%

\begin{figure}
    \centering
		
    \begin{subfigure}[b]{0.35\textwidth}
        \includegraphics[width=70mm,scale=0.75,angle=-0]{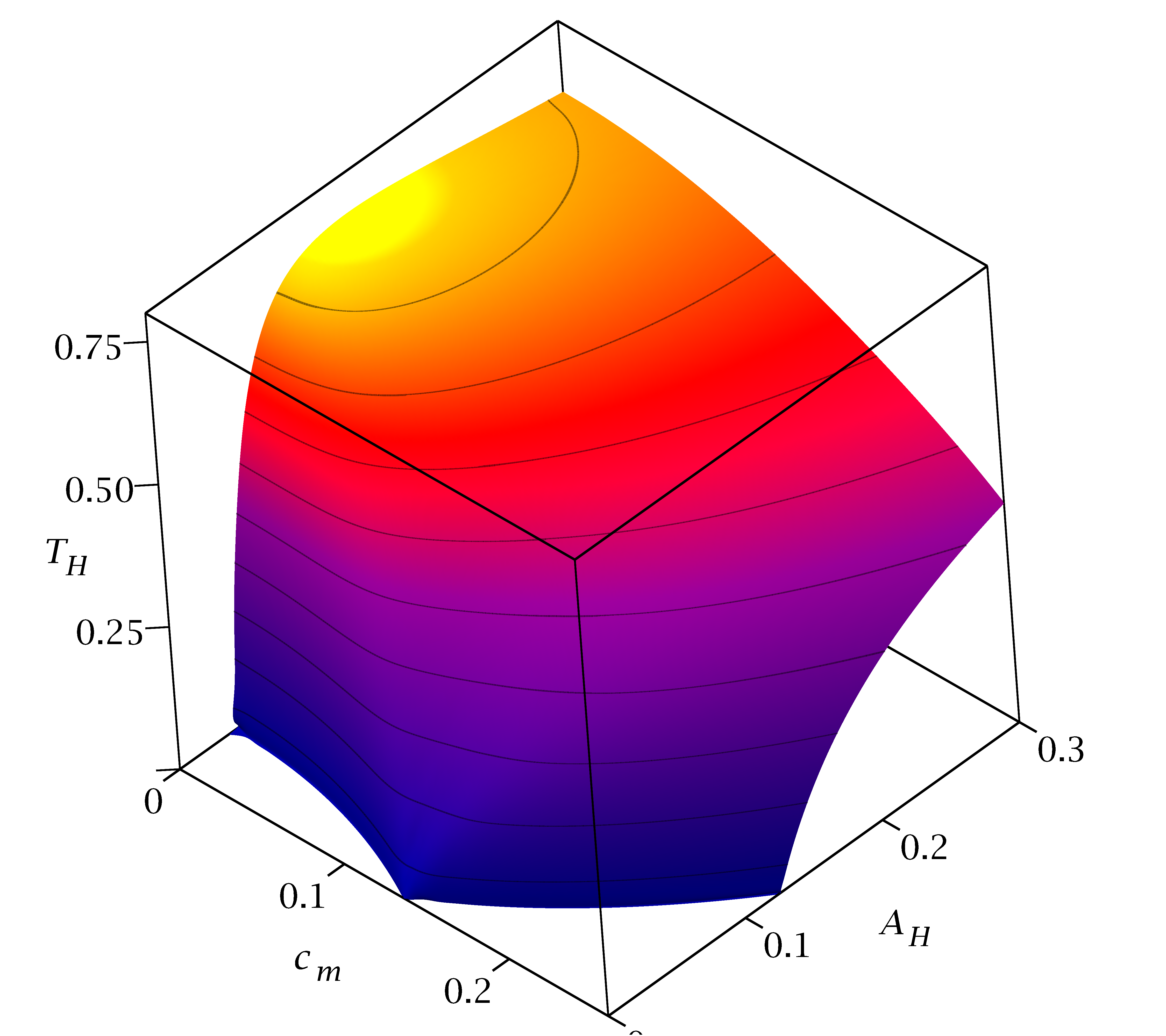}
        %\label{fig:ak_excitations}
    \end{subfigure}~~~~
    \begin{subfigure}[b]{0.35\textwidth}
        \includegraphics[width=70mm,scale=0.75,angle=-0]{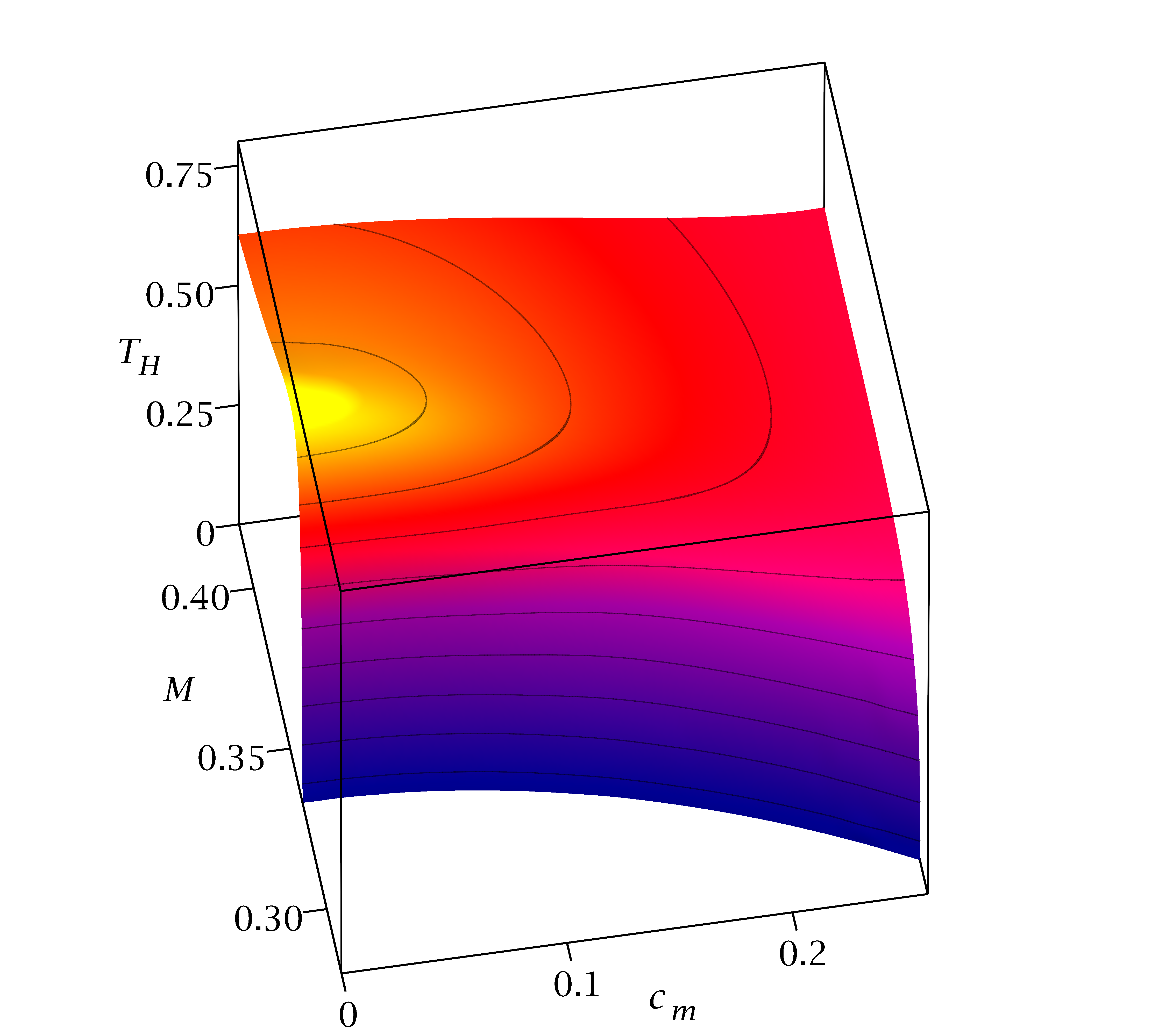}
        %\label{fig:omega_excitations}
    \end{subfigure}
		\\
     \begin{subfigure}[b]{0.35\textwidth}
        \includegraphics[width=70mm,scale=0.75,angle=-0]{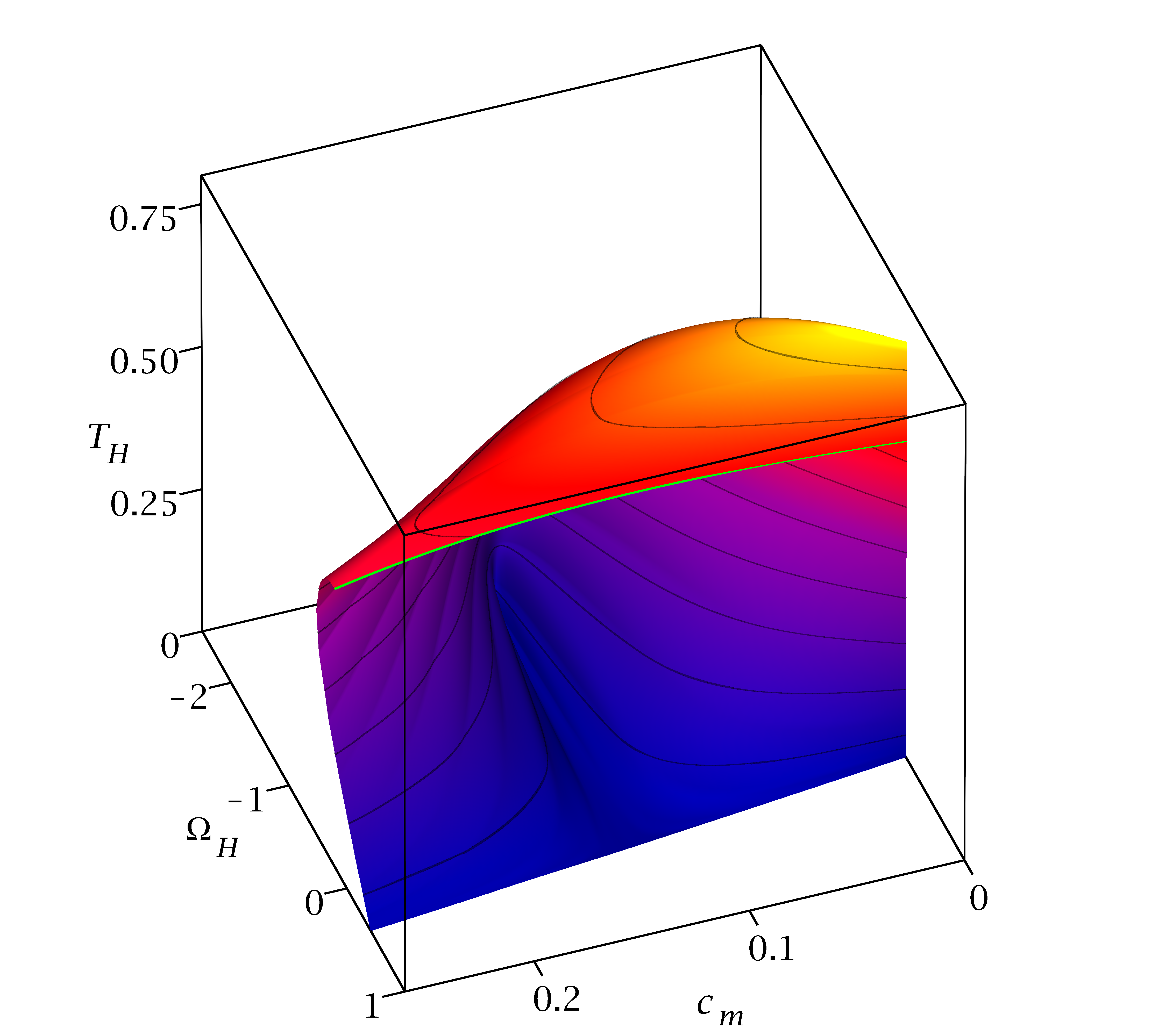}
        %\label{fig:ak_excitations}
    \end{subfigure}~~~~
    \begin{subfigure}[b]{0.35\textwidth}
        \includegraphics[width=70mm,scale=0.75,angle=-0]{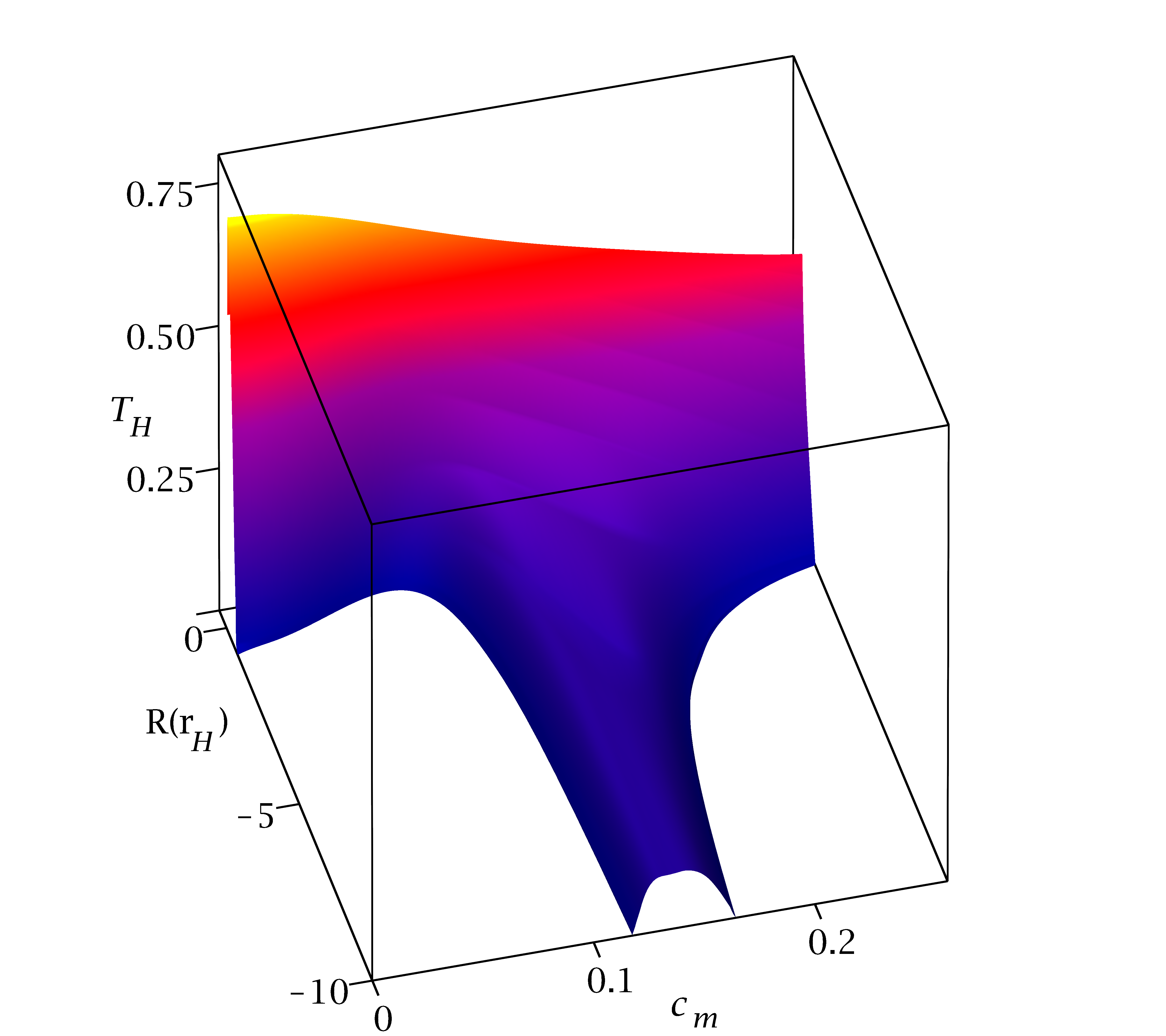}
        %\label{fig:omega_excitations}
    \end{subfigure}
		    \caption{The area $A_H$, mass $M$, angular velocity $\Omega_H$ and Ricci scalar $R(r_H)$ of magnetized black holes 
are shown as a function of $(T_H,c_m)$.
These solutions have $L=1$,
a fixed electric charge $Q^{(R)}=-0.044$ 
and a vanishing angular momentum, $J=0$. }
		\label{BH-prop2}		
\end{figure}

More unusual features occur as well.
For example, in contrast with the CLP case,
one finds BHs which have $J=0$
but still rotate in the bulk\footnote{This feature has been noticed in
\cite{Blazquez-Salcedo:2016rkj}
for $c_m=0$ BHs in EMCS
theory with $\lambda >1$.}.
Some results in this case are shown in Figure \ref{BH-prop2}
for solutions with a fixed value of the electric charge $Q^{(R)}=-0.044$.
These 3D plots exhibit the temperature as a function of $c_m$
and horizon area, mass and horizon angular velocity, respectively.
%To show that these BHs possess a regular horizon, 
The $(T_H;c_m,R(r_H))$-diagram is also included there
(with $R(r_H)$ the Ricci scalar evaluated at the horizon),
to show that these BHs possess a regular horizon\footnote{The
only exception is the
 extremal configurations with $A_H=0$, for
which the Ricci scalar diverges at the horizon at a particular value of $c_m$, where the area vanishes.
% (i.e. the solutions approach a singular planar limit). 
Note that this configuration
marks the separation of two different branches of extremal BHs
and have $\epsilon=0$.
}.
 
%%%%%%%%%%%%%%%%%%%%%%%%%%%%%%%%%%%%%%%%%%%%%%
%%%%%%%%%%%%%%%%%%%%%%%%%%%%%%%%%%%%%%

%%%%%%%%%%%%%%%%%%%%%%%%%%%%%%%%%%%%%%%%%%%%%%%%%%%%%%%%%%%%%%%%%%%%%%%%%%%%%%
\section{Further remarks}
%%%%%%%%%%%%%%%%%%%%%%%%%%%%%%%%%%%%%%%%%%%%%%%%%%%%%%%%%%%%%%%%%%%%%%%%%%%%%%

The main purpose of this paper was to report a generalization of the known
Cveti\v c, L\"u and Pope (CLP) BH solutions 	\cite{Cvetic:2004hs}  
	of the $D=5$ minimal gauged supergravity,
	which contains an extra-parameter in addition to
	mass, electric charge and angular momentum.
This extra-parameter
%called $c_m$ in this work, 
can be identified with the magnitude of the magnetic potential at infinity\footnote{Solutions 
of the minimal gauged supergravity
with a non-vanishing magnetic field on the boundary
have been considered  in \cite{D'Hoker:2009bc}.
However, those solutions possess a Ricci flat horizon, being
asymptotic to Poincar\'e AdS$_5$,
and have very different properties as compared to the BHs in this work.
}.
As such, the solutions here can be viewed as the simplest
AdS$_5$ generalizations of the $D=4$ Einstein-Maxwell solitons and BHs 
recently reported in the literature 
\cite{Herdeiro:2015vaa}-\cite{Herdeiro:2016plq}.
Thus one can predict the existence of a variety of other 
 $D=5$ solutions with the U(1) potentials 
satisfying non-standard far field boundary 
conditions.

The most interesting new feature as compared to the CLP case  
is perhaps the existence of a one parameter family of globally regular, 
smooth solitonic configurations. 
Different from the previously known EMCS solitons with $c_m=0$ which are
supported by the nontrivial topology of spacetime \cite{Cvetic:2005zi},
the solutions here can be considered as deformations of the globally AdS background
and require  a non-vanishing magnetic field on the boundary. 
We also remark that both the BHs and the solitons can be uplifted to type IIB
or to eleven-dimensional supergravity by using the standard results in the literature
(see $e.g.$
\cite{Chamblin:1999tk},
\cite{Gauntlett:2004zh},
\cite{Gauntlett:2007ma}).

\medskip

The study of these magnetized solutions in an AdS/CFT context is an interesting open question.
For example, the background metric upon which the dual field theory
resides is a $D=4$ static Einstein universe with a line element
%\begin{eqnarray}
$
ds^2 =\gamma_{ab}dx^a dx^b=-dt^2+ \frac{1}{4}L^2  (\sigma_1^2+\sigma_2^2 + \sigma_3^2).
$
However, different from the solution in \cite{Cvetic:2004hs},
in this case the theory is formulated in a background $U(1)$ gauge field,
with ${\rm F}_{(0)}=\frac{1}{2}c_m\sigma_2 \wedge \sigma_1$. 
The expectation value  
of the stress tensor of the dual theory 
can be computed by using the AdS/CFT “dictionary”, 
with
$\sqrt{-\gamma} \gamma^{ab}<\tau_{bc}>=$$\lim_{r\to \infty}\sqrt{-h}h^{ab}{\rm T}_{bc}$.

The nonvanishing components of $<\tau_{ab}>$ are
\begin{eqnarray}
&&
\nonumber
 <\tau_\theta^\theta>=<\tau_\phi^\phi >=
\frac{1}{8\pi L}
\left(
 \frac{1}{8}-\frac{5(\hat\alpha-\hat\beta)}{2L^4}-\frac{32 c_m^2}{15L^2}
\right),~
<\tau_\psi^\psi >=
\frac{1}{8\pi L}
\left(
 \frac{1}{8}-\frac{7\hat\alpha-11\hat\beta }{2L^4}+\frac{2c_m^2}{ 5L^2}
\right),~~
%
%\frac{1}{8\pi L}
%\left(
% \frac{2(3\alpha-4\beta)}{ L^4}+\frac{38}{15L^2}
%\right)\cos\theta,~
\\
\nonumber
&&
<\tau_\phi^\psi >= \cos\theta \big (<\tau_\psi^\psi > -<\tau_\phi^\phi > \big),~
<\tau_\psi^t >=\frac{1}{\cos \theta}<\tau_\phi^t >=-\frac{1}{4}L^2<\tau_t^\psi >
=\frac{\hat J}{8\pi L^3}  , 
\\
\nonumber
&&
<\tau_t^t >=
\frac{1}{8\pi L}
\left(
 -\frac{3}{8}+\frac{3\hat\alpha+\hat\beta }{2L^4}-\frac{2c_m^2}{ 15L^2}
\right).
\end{eqnarray}
The trace of this tensor is nonzero, with
\begin{eqnarray}
<\tau_a^a >=-\frac{c_m^2}{2\pi L^3}=-\frac{L}{64\pi}{\rm F}_{(0)}^2,
\end{eqnarray}
resulting from the coupling of the dual theory to a background gauge field \cite{Taylor:2000xw}.
An interesting question here concerns the possible existence,
within the proposed framework, of configurations
possessing a Killing spinor.
However, the results in 
\cite{Cassani:2014zwa}
show that this is not the case:
a supersymmetric solution with a nonzero boundary magnetic field 
is not compatible with the far field  asymptotics (\ref{inf}),
requiring a $squashed$ $S^3$ sphere at infinity.

\medskip 

As avenues for future research,
we remark that the framework and the preliminary results proposed in this work
 may provide a fertile ground for the further study of charged rotating configurations
in the $D=5$ gauged supergravity model.
For example, it would be interesting to study in a systematic way
 their domain of existence, together with the extremal limit.
Moreover, one expects some of the solutions' properties to be generic
when adding scalars or taking unequal spins. 
We hope to return elsewhere with a discussion of some of these aspects.

\bigskip
%%%%%%%%%%%%%%%%%%%%%%%%%%%%%%%%%%%%%%%%%%%%%%%%%%%%%%%%%%%%%%%%%%%%%%%
{\bf Acknowledgement}
%%%%%%%%%%%%%%%%%%%%%%%%%%%%%%%%%%%%%%%%%%%%%%%%%%%%%%%%%%%%%%%%%%%%%%%
\\
We gratefully acknowledge support by
the DFG Research Training Group 1620 ``Models of Gravity''
and by the Spanish Ministerio de Ciencia e Innovaci\'on,
research project FIS2011-28013.
 E. R. acknowledges funding from the FCT-IF programme.
This work was also partially supported 
by  the  H2020-MSCA-RISE-2015 Grant No.  StronGrHEP-690904, 
and by the CIDMA project UID/MAT/04106/2013.  
J.L.B.S. and J.K. gratefully acknowledge support by the grant FP7, Marie Curie Actions, People, International Research Staff Exchange Scheme (IRSES-606096). F. N.-L. acknowledges funding from Complutense University under Project No. PR26/16-20312.

%%%%%%%%%%%%%%%%%%%%%%%%%%%%%%%%%%%%%%%%%%%%%%%%%%%%%%%%%%%%%%%%%%%

\end{document}